# Authoring and Living Next-Generation Location-Based Experiences

Olivier Balet[1]
DIGINEXT

Boriana Koleva
University of Nottingham

Jens Grubert
Graz University of Technology

Kwang Moo Yi
Ecole Polytechnique Fédérale de Lausanne

Marco Gunia
Technische Universität Dresden

Angelos Katsis
EXUS

Julien Castet
Immersion

**ABSTRACT**

Authoring location-based experiences involving multiple participants, collaborating or competing in both indoor and outdoor mixed realities, is extremely complex and bound to serious technical challenges. In this work, we present the first results of the MAGELLAN European project and how these greatly simplify this creative process using novel authoring, augmented reality (AR) and indoor geolocalisation techniques.

**Index Terms:** H.5.1 [Multimedia Information System]: Artificial, augmented, and virtual realities

## 1 INTRODUCTION

The latest wave of innovation for computer games is *mobile* and, more precisely, *location-based*. These pervasive experiences [1] radically differ from traditional computer games or their mobile equivalent. They focus the players' attention on the real world as much as on the digital world of the game, aiming to merge the two in the form of mixed realities. Although location-based gaming is an industry on the verge of explosive growth, the creation and deployment of such experiences, especially those involving multiple participants, is simply out of reach for the vast majority of creative industries and authors because of the many state-of-the-art technologies they require, their hard to master limitations, and the complex concepts they employ.

Thirteen European organizations partnered in 2014 in the frame of the MAGELLAN research project to address these issues, adopting a holistic approach to implement an integrated European platform for both authors and participants. This project first aims to deliver an authoring environment based on visual authoring and natural user interface principles to enable non-programmers to author and publish multi-participant location-based experiences. Its second objective is to deliver a scalable web platform supporting the sharing, browsing and execution of a massive number of such experiences. Finally, MAGELLAN will produce a series of guides for authors of location-based experiences that will constitute a reference for future research in the field.

## 2 OVERALL ARCHITECTURE

MAGELLAN implements a novel and scalable Web platform enabling the collaborative authoring and publication of experiences as well as the participation of numerous users in these experiences. This platform (fig. 1) is based on an open framework that provides required standard technologies to store user accounts, assets, and that is responsible for the execution of experiences. It is designed to be scalable using a big-data storage and cloud-computing architecture to support a massive number of users and ongoing experiences.

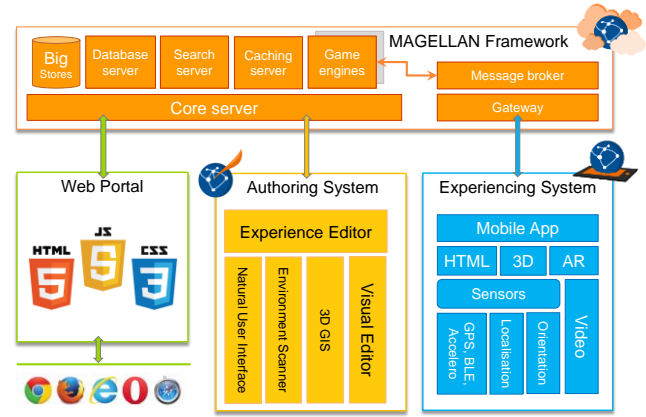

Figure 1: MAGELLAN overall architecture.

## 3 AUTHORING SYSTEM

The Authoring system aims to make the creation process accessible to laymen and non-programmers. It is built on INSCAPE, a visual interactive storytelling tool developed in the frame of the eponym European project and extended in the CHESS FP7 project [2]. MAGELLAN is further extending this system to support the authoring of interactive multi-participants experiences taking place in AR environments. For that purpose, the 3D underlying engine has been redesigned to support geodetic environments and geographic information streamed from standardized sources (OGC compliant). It is connected to the Web platform to exploit assets shared by other authors (e.g. 3D elements, scenarios) or to publish assets and experiences. An experience is modeled as a graph of Activities (e.g. quizzes, 3D or AR games), that participants can perform according to their position in the real world, interactions with other participants or with the virtual entities populating the experience. Pre-conditions and controllers are used to control the unfolding of the scenario.

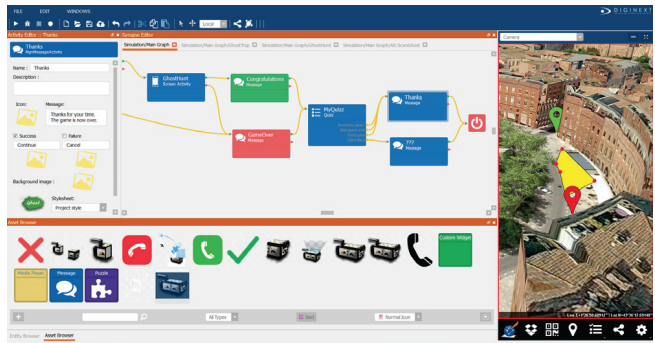

Figure 2: The MAGELLAN visual storytelling tool

---

[1] forename.name@diginext.fr

The MAGELLAN project also implements a natural user interface (fig. 3) featuring tangible objects that can be used in combination with interactive surfaces to control the authoring tool and make the creative process even more intuitive. We also introduced an original method that enables the author to easily create customized tangible objects with a 3D printer.

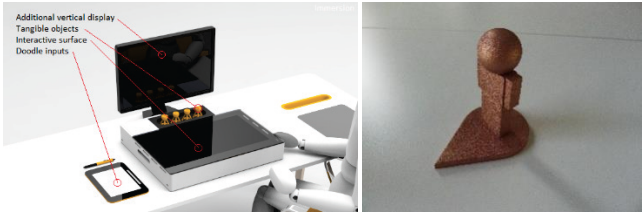

Figure 3: The MAGELLAN natural user interface

## 4 EXPERIENCING MOBILE SYSTEM

The MAGELLAN mobile app runs on the participant's smartphone connected to the Web platform. We implement a number of innovative features to cope with current geolocalisation and mobile AR limitations.

### 4.1 Indoor localization

A seamless hybrid indoor and outdoor positioning system is being developed combining multiple standards to enable the best possible localization depending on the individual conditions. Amongst others, GPS, WLAN, and Frequency-modulated continuous-wave (FMCW) radar are incorporated. For indoor scenarios, FMCW radar provides a highly accurate technique with a large coverage range. This is to be complemented by WLAN, based on the received signal strength. Combined, both provide an optimal balance between accuracy and costs to fit user needs. So far, a basic approach based on GPS and FMCW-radar has been implemented [3], which is to be enhanced by further standards in later versions. The mobile unit is capable of determining its own position using the information from the reference stations. Furthermore, a study was performed, which shows the potential to enable 1 cm position accuracy and a reduction of power consumption by 50% compared to current FMCW implementations.

### 4.2 3D Tracking

We implemented a panorama-based AR tracking component that works on mobile platforms under a variety of lighting conditions. To this end, we developed a new machine learning based feature point detector that outperforms state-of-the-art methods. It is robust to illumination changes and exhibits high repeatability in outdoor scenes taken on different time of day and seasons. We can reliably extract salient features and compute the geometric transform between a panorama built with the authoring tool and a query image, such as one taken by a smartphone (fig. 4, right). For comparison, SIFT feature points, used as a baseline, fails in cases such as the one depicted in figure 4, left. By contrast, as shown in figure 4, using our own features produces a correct result.

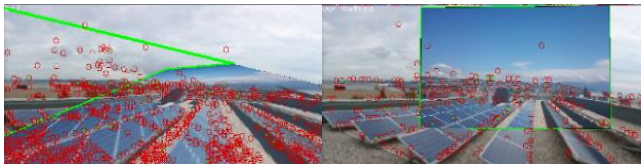

Figure 4: Image Matching with SIFT (left), our approach (right).

### 4.3 Context-Aware Augmented Reality

Understanding how context sources influence the utility of AR is crucial for designing MAGELLAN experiences as they take place in changing mobile contexts. We are working on a better understanding of those context sources and how to better support the uptake of AR in public space. For example we introduced a lightweight approach for augmenting public displays without the need for extensive infrastructure [4]. This should facilitate the distribution of MAGELLAN experiences in public space as public displays provide ample opportunities for visible entry points to those experiences. Our approach is based on multicasting display content through real-time streaming protocols to nearby clients. It scales easily to many clients and supports user-perspective rendering of display content on clients with front facing cameras and dual camera access (fig. 5). In the near future, we want to explore how to combine AR with alternative interfaces for seamless interaction on and away from physical artifacts, going beyond simple shapes like posters [5].

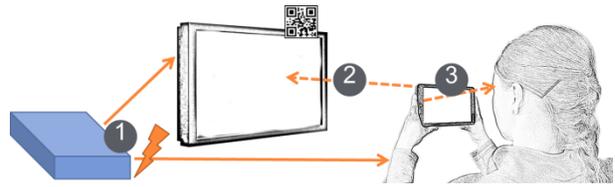

Figure 5: A public display content source streams screen content to mobile clients (1) that track the public display via natural feature tracking (2) and the users face (3) to enable user perspective AR.

## 5 CONCLUSION

We believe that the results of the research performed in the frame of MAGELLAN can have a significant impact on a wide range of location-based and mobile AR applications. While the system currently exists as a proof-of-concept, it has already demonstrated effectiveness enabling non-programmers to create and publish, in a few hours, complex experiences featuring VR and AR activities, which would have required multiple technical expertises and extensive efforts to implement with current approaches. The coming months will see the release of the beta version of the system and the availability of the scientific results of this European research initiative to a wider audience.

### ACKNOWLEDGEMENTS

This project has received funding from the European Union's 7th Framework Programme for research, technological development and demonstration under grant agreement no ICT-FP7-611526.